\documentclass[aps, prl, twocolumn, showpacs, floatfix,10pt,superscriptaddress]{revtex4-2}
\usepackage{amsmath}
\usepackage{amsfonts}
\usepackage{amsbsy}
\usepackage{amssymb}
\usepackage{graphicx}
\usepackage{textcomp}
\usepackage[caption=false,singlelinecheck=false]{subfig}
\usepackage{xcolor}
\usepackage{mathrsfs}
\usepackage{mathtools}
\usepackage{bm}
\usepackage{gensymb}
\usepackage{braket,wasysym}
\usepackage[colorlinks,linkcolor=blue,citecolor=red,filecolor=magenta,urlcolor=red,breaklinks]{hyperref}
\usepackage[bottom]{footmisc}
\usepackage{verbatim}

\hypersetup{colorlinks=true, urlcolor=blue, citecolor=red, pdfborder={0 0 0}}
\usepackage{breakurl}
\usepackage{natbib}

\allowdisplaybreaks

\newcommand{\tcm}[1]{\textcolor{magenta}{#1}}

\normalfont
\def\be{\begin{equation}}
	\def\ee{\end{equation}}
\def\bea{\begin{eqnarray}}
	\def\eea{\end{eqnarray}}

\begin{document}
\title{Hidden hyperspace geometry and long-distance quantum coupling}

\author{Junmo Jeon}
\email{junmo1996@kaist.ac.kr}
\affiliation{Korea Advanced Institute of Science and  Technology, Daejeon 34141, South Korea}
\author{SungBin Lee}
\email{sungbin@kaist.ac.kr}
\affiliation{Korea Advanced Institute of Science and  Technology, Daejeon 34141, South Korea}

\date{\today}
\begin{abstract}
Most periodic systems are governed by short-range interactions as long-range interactions in these systems diminish uniformly.
In this letter, however, we demonstrate that this is not true for a more general class of systems, which possess long-range order without periodicity, known as quasiperiodic systems.
Quasiperiodicity alters the well-known characteristics of the long-range couplings, resulting in anomalous enhancement for arbitrarily long distances, even beyond the mesoscopic scale. By exemplifying the indirect spin exchange interaction, we show that the long-range coupling in a quasiperiodic chain does not attenuate over distance but is instead governed by a novel distance metric, we have named the hyperspace geometric distance. This enables us to remotely control the spins over even mesoscopic distances.
Our work provides new paradigms of strongly correlated physics applicable to a broader class of systems beyond the conventional ones.
\end{abstract}
\maketitle

\textit{\tcm{Introduction---}}
Exploring long-range quantum correlation opens up a rich landscape of intriguing physical phenomena and applications across various fields of study including quantum computing and material science\cite{kulik2012quantum,teh2023mesoscopic,imry1998mesoscopic,frowis2018macroscopic,kurizki2015quantum,leggett2004quantum}. 
The entanglement emergent beyond mesoscopic scales has garnered attention from both theoretical perspectives and the quantum information engineering due to their high controllability \cite{shevchenko2019mesoscopic,iachello2004quantum,sellitto2016mesoscopic,datta1997electronic}.
However, we face significant challenges in realizing their potential. Notably, the generation and manipulation of entanglement at a distance longer than mesoscopic scales remain a poorly understood field despite its central role in quantum information science\cite{gallego2021macroscopically,stav2018quantum,wilde2013quantum,leggett2004quantum,raimond2001manipulating}.
Since most of the systems are governed by short-range interactions, the quantum correlations decay rapidly with distance\cite{chaikin1995principles,de2006entanglement}. Even in the case of long-range interactions, the strength of the interaction decays with distance according to a power law\cite{cahaya2022adiabatic,PhysRevB.73.214205,yosida1996theory,miyazaki2020magnetic,PhysRevLett.111.196601}. This makes hard to utilize the long distance quantum entanglement beyond the mesoscopic scales by using conventional interactions in crystals. In this regards, one could ask how long-range interactions would be characterized if the system is aperiodic involving extra-dimensional degree of freedom.

The systems involving hidden higher dimensional structure are called hypermaterials such as quasicrystals\cite{watanabe2021theory}. Their inherent higher dimensional structure is referred to as the hyperspace which enables us to define the geometrical properties of hypermaterials based on their extra-dimensional space known as perpendicular space. In the hypermaterials, the sites characterized by the proximate locations in the perpendicular space, share the local surroundings across a broad region, and hence many physical characteristics are shared between them regardless of their actual physical distance. For example, the critical state neither exponentially localized nor extended is concentrated at the distant sites characterized by the proximate locations in the perpendicular space\cite{steurer2011quasicrystals,jagannathan2021fibonacci}.
Hence, the distances within the perpendicular space, are crucial in shaping the quantum mechanical properties of hypermaterials, overriding the significance of physical distances. Particularly, the proximate locations in the perpendicular space can strongly interact despite their long distance. This leads to the anomalous long-distance coupling characterized by the hyperspace.

\begin{figure}
  \includegraphics[width=0.5\textwidth]{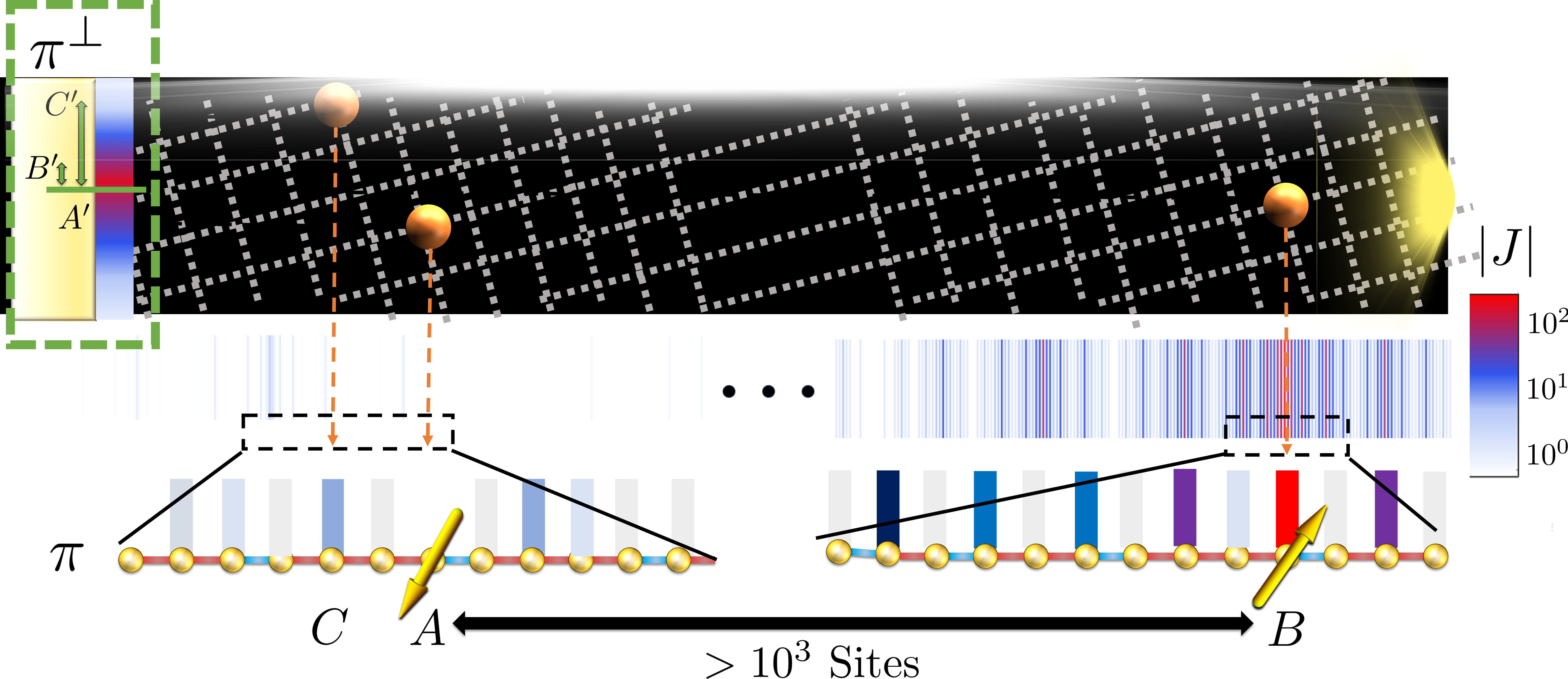}
  \caption{Schematic illustration and numerical calculation of the long-range interaction in the quasiperiodic chain generated by cut-and-project scheme. The quasiperiodicity is originated from the irrational slope of the physical space ($\pi$) with respect to the 2D square lattice represented by the gray dashed lines. There is an one-to-one correspondence between the sites of the quasiperiodic chain and the hyperlattice sites, and  their projections on the perpendicular space ($\pi^\perp$) drawn as the yellow bar. Quasiperiodic pattern consists of two (or more) nearest neighbor hopping represented by red and blue bars. Yellow spheres are the lattice sites. The peaks (white-blue-red) show the spatial distribution of the magnitude of long-range interaction, $J$ for $A$, where the local magnetic impurity (yellow arrow) exists. $\vert J\vert$ are the relative values calculated for the silver mean tiling. The spin at $B$ more than $10^3$ times that of lattice sites far apart from the $A$, strongly interacts with the spin at the $A$ compared to the neighboring site, for instance $C$. Such strongly interacting sites share the local environment. Thus, their distance within $\pi^\perp$ is very short as shown $A'$ and $B'$. The coupling strength decays as a function of distance within the perpendicular space as emphasized by the green arrows.}
  \label{fig: scheme}
\end{figure}

In this letter, we study the anomalous properties of the long-range exchange interaction of the localized moments in one-dimensional quasicrystals mediated by the critical electronic states. By introducing the hyperspace geometric distance, we show that the long-range couplings in the quasicrystals could be characterized in terms of it rather than the physical distance. For concrete argument, we consider the Fibonacci quasicrystal\cite{kellendonk2015mathematics}. Focusing on the half-filled case, we point out two interesting features of the long-range interaction in the quasiperiodic chains. First, unlike the case of periodic crystals, the interaction does not universal decay in physical space, but is exceptionally enhanced between the sites far apart. Second, the strongly interacting sites share the broad environment, and hence their hyperspace geometric distance is small. Fig.\ref{fig: scheme} illustrates such anomalous interaction in quasicrystal, where two spins, marked as yellow arrows at $A$ site and $B$ site, can strongly interact with each other even over long-distances more than $10^3$ sites, a beyond mesoscopic length scale. On the other hand, in the perpendicular space, they are in neighbor as shown $A'$ and $B'$. Consequently, the spins form the strongly entangled pair over long-distances. It opens up the capability of manipulating spins (qubits) in a non-local way via quasiperiodic patterns, that reveals long-range quantum entanglement.

\textit{\tcm{Hyperspace geometric distance---}}
Let us consider the nearest neighbor tight binding model together with localized spins in 1D chain whose Hamiltonian is given by
\begin{align}
\label{H}
&H=-\sum_{\braket{i,j}}t_{i,j}(c_i^\dagger c_j+c_j^\dagger c_i)+J_K\sum_i\vec{S}_i\cdot\vec{s}_i.
\end{align}
Here, $t_{i,j}$ is the nearest neighbor hopping parameters. $\vec{S}_i$ is the localized spin at the $i$ site, and $\vec{s}_i=c_i^\dagger\vec{\sigma}c_i/2$ is the electron's spin at the $i$ site, where $c_i$ is the annihilation operator for the electron at the $i$ site, $\vec{\sigma}$ is the Pauli matrices. The site index runs from 1 to $N$. Although we will consider $J_K$, the Kondo coupling in this work, we emphasize that our work could be applicable for general local qubits interactions such as multipole-orbital interaction\cite{nejati2017kondo,gulacsi2004one,minami2015low,jeon2024unveiling}. Under the open boundary condistion, the tight binding Hamiltonian in Eq.\eqref{H} has the sublattice symmetry. Thus, the electronic spectrum is symmetric with respect to the zero energy, regardless of the pattern of $t_{i,j}$\cite{PhysRevResearch.3.013168}. We focus on the half-filled case where the Fermi level is placed exactly at the zero energy. By integrating out itinerant electrons, the Hamiltonian becomes $\mathcal{H}=\sum_{i\neq j}J_{ij}\vec{S}_i\cdot\vec{S}_j$ where $J_{ij}$ is the long-range coupling between $i$ and $j$ sites given by,
\begin{align}
\label{RKKY2}
J_{ij}=\frac{J_K^2}{4}\sum_{m,n,E_m\neq E_n}&\mathcal{R}[\psi_m(i)\psi_m(j)^*\psi_n(j)\psi_n(i)^*] \\ &\times \frac{n_F(E_n-E_F)-n_F(E_m-E_F)}{E_n-E_m}.\nonumber
\end{align}
Here, $\hbar=1$, $E_F$ is the Fermi energy. $\mathcal{R}[x]$ is real part of $x$, respectively. $n,m$ are indices of the eigenstates, and $\psi_{n}(i)=\langle i \vert n\rangle$ is the wave function of the energy eigenstate $\vert{n}\rangle$ whose energy is $E_n$ at the $i$ site. $n_F(x)=(1+\exp(\beta x))^{-1}$ is the Fermi distribution function, where $\beta=1/\tau$ and $\tau$ is temperature. We focus on the zero temperature limit. Note the states near the Fermi level mainly contribute to the interaction\cite{zhou2010strength}. The interaction in Eq.\eqref{RKKY2} in the periodic systems exhibit universal decay as $J_{ij}\sim R_{ij}^{-1}$, where $R_{ij}$ is the distance between $i$ and $j$ sites\cite{PhysRevB.36.3948,PhysRevB.58.3584}. Thus, it is negligibly small over a long distance. However, as we will show, this does not hold in quasiperiodic systems, where the electron states are neither localized nor extended but critical.


Now let us consider the quasiperiodic chains generated by the cut-and-project scheme. On the 2D square grid points called by the hyperlattice, we consider two orthogonal 1D subspaces called by the physical space ($\pi$) and perpendicular space ($\pi^\perp$), respectively (See Fig.\ref{fig: scheme}). They take the irrational slope with respect to the hyperlattice plane. Thus, by projecting the hyperlattice to $\pi$, we get the quasiperiodic structure. Here, we project the hyperlattice point only if its projection image onto $\pi^\perp$ is included in the compact window, which is given by the projection of the Wigner-Seitz cell of the hyperlattice\cite{senechal1996quasicrystals}. For instance, the Fibonacci quasicrystal is generated by projecting the hyperlattice to the half-line $y=x/\phi+(1-\phi/2)$ starting from $(4-3\phi,3-\phi)/10$, where $\phi$ is the golden ratio\cite{kellendonk2015mathematics}.

The Fibonacci quasicrystal is comprised of two different nearest neighbor distances. Hence, it consists of two prototiles, say A and B assigned different hopping integrals $t_\mathrm{A}$ and $t_\mathrm{B}$, respectively. The strength of the quasiperiodicity is given by the ratio between the two hopping parameters, $\rho \equiv t_\mathrm{A}/t_\mathrm{B}$.
The sublattice symmetry and gap labeling theorem\cite{bellissard1992gap,bellissard2006spaces,kellendonk2015mathematics} guarantee the presence of the zero energy state in the Fibonacci quasicrystal, which is inductively given by $\frac{\psi(2k+1)}{\psi(1)}=(-1)^k \rho^{h(k)}$\cite{PhysRevB.96.045138}. Here, $h(k)$ is called by the height field, which is the pattern-dependent function given by
\begin{align}
\label{height}
h(k)=\sum_{0\le i<k}\mbox{sgn}\left(\frac{\rho-1}{t_\mathrm{A}}(t_{2i+1,2i+2}-t_{2i+2,2i+3})\right),
\end{align}
where $t_{2i+1,2i+2}$ is either $t_\mathrm{A}$ or $t_\mathrm{B}$ depending on the prototile A and B between the site ${2i+1}$ and  site ${2i+2}$\cite{PhysRevB.96.045138,PhysRevResearch.3.013168}. Note that the zero energy state is critical concentrated on a few distant sites sharing the local patterns\cite{PhysRevB.96.045138,PhysRevResearch.3.013168,PhysRevB.106.134431}. Besides, in the Fibonacci quasicrystal, there are many critical states near zero energy\cite{jagannathan2021fibonacci}. Thus, these critical states mainly contribute to the long-range interaction for $E_F=0$.

The projection maps onto $\pi$ and $\pi^\perp$ are injective due to the irrational slope. Thus, the positions within these subspaces are compatible. Importantly, close locations within $\pi^\perp$ imply matching patterns of the wider environment. Thus, the critical states are localized within the window\cite{PhysRevResearch.3.013168,jagannathan2021fibonacci}. Consequently, as we will show, there would be a strong correlation between the sites whose $\pi^\perp$ positions are adjacent but physically distant. Therefore, it is important to introduce the distance within the perpendicular space rather than the physical space---which we will call the hyperspace geometric distance.

The symmetric projection images on $\pi^\perp$, say $X$ and $-X$ correspond to palindromic pair of the local patterns, such as AB and BA. Since the Hamiltonian does not distinguish them, we should introduce the hyperspace geometric distance within $\pi^\perp$ for $x_1$ and $x_2$ sites in $\pi$, say $d_{\pi^\perp}(x_1,x_2)$ as
\begin{align}
\label{distance}
&d_{\pi^\perp}(x_1,x_2)=\vert \vert\pi^\perp(x_1)\vert-\vert\pi^\perp(x_2)\vert \vert,
\end{align}
where $\pi^\perp(x_1)$ is the projection image on $\pi^\perp$ corresponding to $x_1$.


In the following, we will show that the critical states, which are highly concentrated on a few sites give rise to the anomalous long-range coupling in the Fibonacci quasicrystal. Thus, one needs to define the geometric distance of the position on the perpendicular space with respect to the set of reference points, say $M$ where the relevant critical state is highly concentrated on. Given $E_F$, we define $M=\{x\vert \vert\psi_{\varepsilon_F}(x)\vert\ge\vert\psi_{\varepsilon_F}(y)\vert \ \ \forall y\in\pi\}$, where $\psi_{\varepsilon_F}$ is the electronic eigenstate whose energy, $\varepsilon_F$ is the closest to $E_F$. Thus, the sites in the set $M$ are the most dominant positions for long-range interaction. For instance, we have $M=\{x \vert h(x)\le h(y) \ \ \forall y\in \pi \}$ for $\rho<1$ and $E_F=0$. For the Fibonacci quasicrystal, $\pi^\perp (M)=\{0\}$. Let us define the hyperspace geometric distance between $M$ and the site $i$, say $d_{\pi^\perp}(i,M)=\inf_{x \in M}d_{\pi^\perp}(i,x)$. Note the choice of $M$ depends on the Fermi level we are considering in, parameters of the Hamiltonian such as $\rho$ and tiling pattern. For given $M$, now we define the hyperspace geometric distance which tells us how two sites would be strongly correlated via the critical state related to $M$, say $D_{\pi^\perp}^{M}(i,j)$ as
\begin{align}
\label{DisDis}
&D_{\pi^\perp}^{M}(i,j)=d_{\pi^\perp}(i,M)+d_{\pi^\perp}(j,M).
\end{align}

\begin{figure}
  \includegraphics[width=0.44\textwidth]{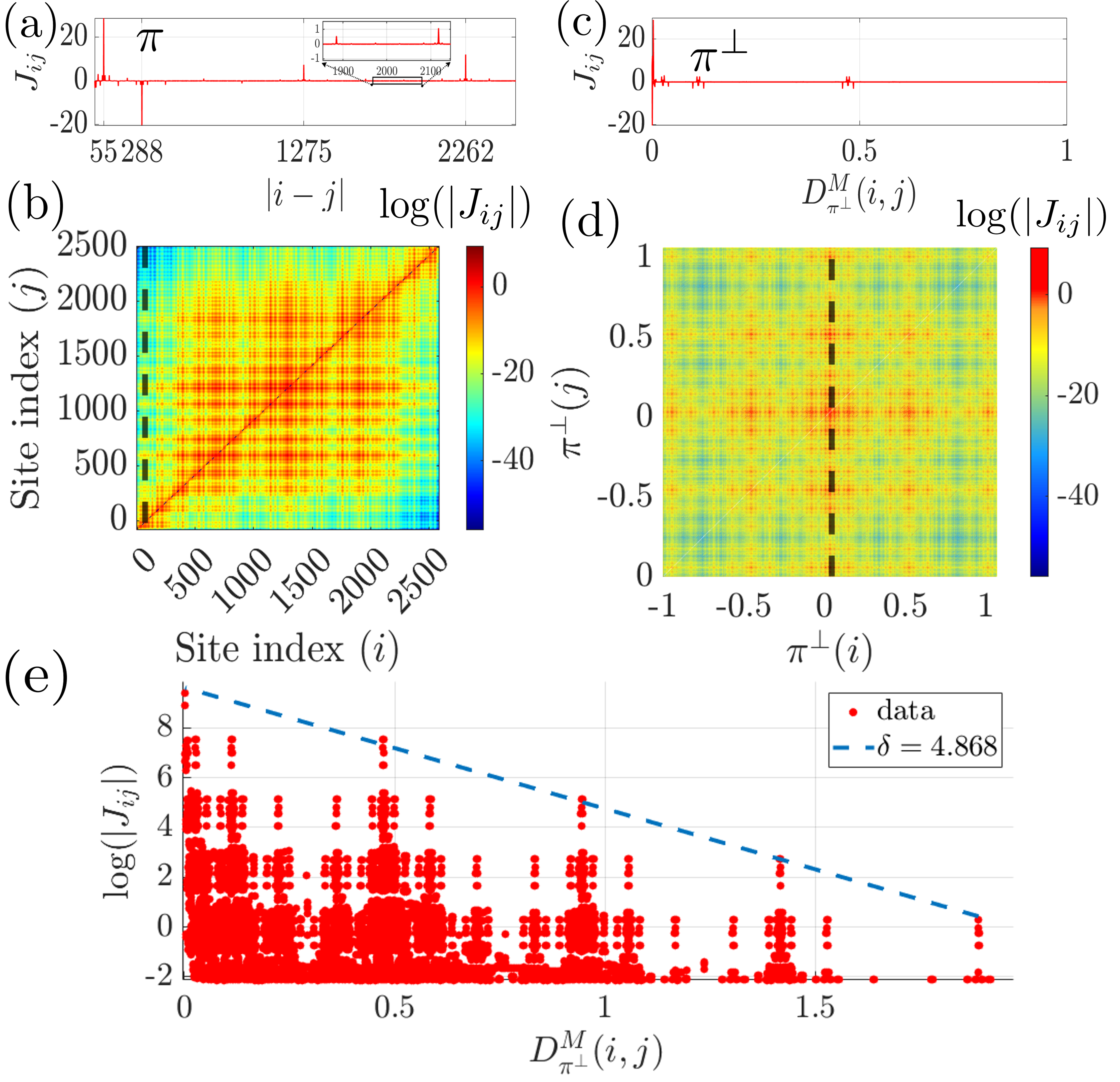}
  \caption{Anomalous indirect interaction in the Fibonacci chain viewing on the (a,b) physical space ($\pi$) and (c,d) perpendicular space ($\pi^\perp$). Coupling $J_{ij}$, in the Fibonacci quasicrystal for $i=17$ as a function of (a) physical distance $\vert i-j\vert$ and (c) hyperspace geometric distance $D_{\pi^\perp}^M(i,j)$, where $\pi^\perp(M)=\{0\}$. The coupling is anomalously enhanced even for the $10^3$ sites distance. The inset of (a) is emphasizing the fractal structure of $J_{ij}$. The dominant couplings mainly emerge at the origin of the perpendicular space. General indirect interaction strength in Fibonacci quasicrystal in terms of (b) the site indices and (d) $\pi^\perp$ projection images, respectively. Off-diagonal bright region in (b) indicates the long-distance strong interaction. Black dashed lines in (b) and (d) represent the location of $i=17$ site and its $\pi^\perp$ projection image, respectively. (e) Exponential decay and fractal behavior of the coupling strength on the perpendicular space. The envelope of the coupling strength exponentially decays as $\vert J_{ij}\vert\sim \exp(-\delta D_{\pi^\perp}^M(i,j))$, where $\delta\approx 4.868$. The system size $N=2585$, the strength of the quasiperiodicity $\rho=0.3$ and $J_K=1$.}
  \label{fig: main1}
\end{figure}

\textit{\tcm{Anomalous long-distance strong coupling---}}
From now on, we discuss the long-range interaction in the quasiperiodic chain. For a concrete argument, we focus on the Fibonacci quasicrystal case. See Supplementary Materials for the general quasiperiodic chain cases. 
Focusing on the half-filled case, we calculate the coupling, $J_{ij}$ as a function of distance $|i-j|$ for $i=17$ in Fig.\ref{fig: main1} (a).

Let us point out important features of the long-range coupling in the Fibonacci chain. First of all, unlike the periodic system, Fig.\ref{fig: main1} (a) shows that $J_{ij}$ is exceptionally enhanced even for the $10^3$ lattice sites scales in the Fibonacci quasicrystal. Since this length scale is comparable with the system size $N$, it could be arbitrarily increased in the thermodynamic limit. This indicates that strong correlation over mesoscopic distances is possible in the Fibonacci chain. Second, $J_{ij}$ show the fractal structure (See the inset of Fig.\ref{fig: main1} (a).). Such fractal behavior of $J_{ij}$ is originated from the self-similar wave functions near the Fermi level. Remarkably, Fig.\ref{fig: main1} (b) demonstrates that such fractal and strong long-distance interactions in the Fibonacci quasicrystal are general characteristics.

Importantly, the sites connected by strong long-distance interactions are not neighboring nor randomly distributed, but instead, they share a common long-range local environment.
To clarify this, we use the hyperspace geometric distance, $D_{\pi^\perp}^M(i,j)$ introduced in Eq.\eqref{DisDis}. 
Fig.\ref{fig: main1} (c) and (d) show the interaction strength as a function of the position of the perpendicular space. Unlike Fig.\ref{fig: main1} (a), where the interaction is enhanced for long distances, Fig. \ref{fig: main1} (c) and (d) show that the dominant couplings exist when $\pi^\perp$ projection images of two interacting sites are placed close to zero and are adjacent to each other. Thus, the strongly coupled sites share the same local tiling pattern.
Fig.\ref{fig: main1} (e) illustrates the exponential decay of the coupling strength, $\vert J_{ij}\vert$ as the function of the hyperspace geometric distance, $D_{\pi^\perp}^M(i,j)$. Along with the fractal structure of $J_{ij}$, the envelope of the coupling strength is uniformly decaying as a function of $D_{\pi^\perp}^M(i,j)$. Specifically, $\vert J_{ij}\vert\sim \exp(-\delta D_{\pi^\perp}^M(i,j))$, where $\delta=4.868$ for $\rho=0.3$. 
Therefore, in the Fibonacci quasicrystal, utilizing hyperspace geometric distance is a more suitable approach for understanding long-range coupling compared to physical distance. It turns out that 
the analyzing the local tiling patterns in relation to the geometry within the perpendicular space plays a pivotal role in elucidating the correlation physics of hypermaterials.


Before discussing the application of strong interactions, we have two notable points about anomalous interactions. First, along with the critical states near the Fermi level, the long-distance strong interaction would be generically emerges by enhancing the concentration of the critical states in terms of $\rho$. This is independent on the specific choice of the quasiperiodic tiling pattern\cite{PhysRevB.60.322}. Second, anomalous long-distance interaction is absent in the approximants. Since the electronic wave functions of the approximant are Bloch states, $J_{ij}$ is universally decaying as the function of physical distance. See Supplementary Materials for detailed information.

\begin{figure}
\centering
  \includegraphics[width=0.48\textwidth]{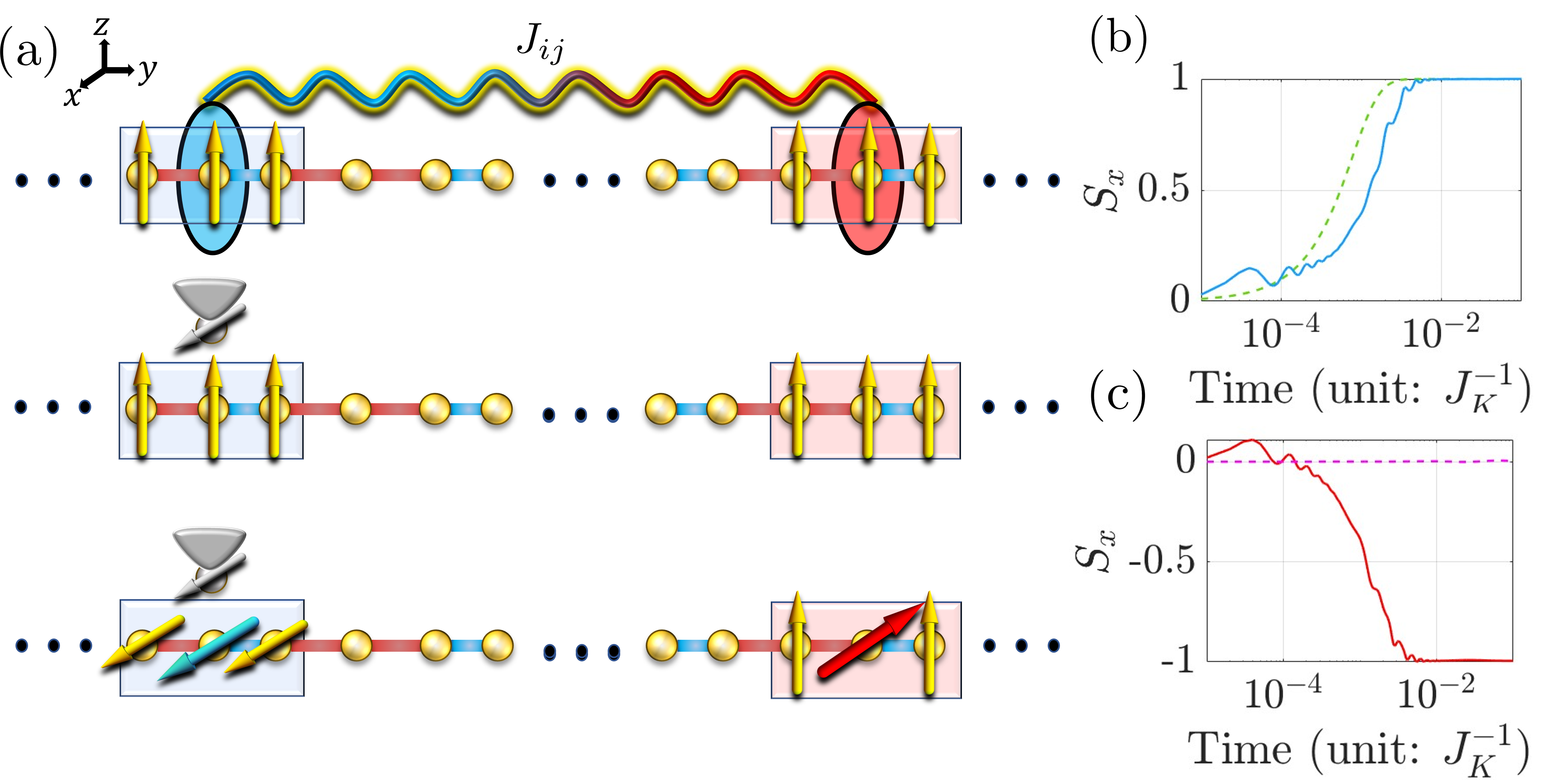}
  \caption{(a) Schematic illustration of a proposed experimental scenario for non-local control of the magnetic moments. Red and blue bars represent two different hopping terms of the Fibonacci chain. (Top-Set up) Place the local magnetic impurities on two separated regions emphasized by blue and red shaded squares. Align the local moments along the $z$-axis initially by applying strong magnetic field, for instance. Here, blue and red ellipses and wiggled curve are drawn for emphasizing a strongly correlated pair via long-distance coupling, $J_{ij}$. (Middle) Apply the local external field along the $x$-axis on the blue region only by using the spin-polarized STM tip (gray). (Bottom) The local moment in the red region  (particularly for the red arrow) is aligned along the $x$-axis due to the strong $J_{ij}$ with the cyan arrow. (b) $S_x$ for the spins under the external field as a function of time. Cyan curve represents for the site $i=17$, while green dashed curve represents for the site $i=16$. (c) $S_x$ for the spins in the region where the local field is absent. Red curve represents $S_x$ as a function of time for the site $i=2279$. This is different from the behavior of the moment at the site $i=2278$ (pink dashed line) which remains along the $z$-axis. $B=10^4$, $N=2585$ and $J_K=1$.
 }
  \label{fig: main2}
\end{figure}

\textit{\tcm{Long-range entangled pair---}}
Considering exceptionally enhanced interactions for a long-distance, one can generate and control the stable long-range entangled pairs. To reduce the complexity, we consider two widely separated local regions, $D$ and $D'$, where magnetic moments are located. These two local regions include sites that are far away, but their indirect interaction is greatly strong. Fig.\ref{fig: main2} illustrates the control of local magnetic moments due to the strong coupling between long-distance spins. Initially, the local magnetic moments are aligned along the $z$-axis, as shown in the first row of Fig.\ref{fig: main2} (a). At time $t=0$, we apply the local field along the $x$-axis, $B\hat{x}$ by using the spin-polarized STM tip (gray tips in Fig.\ref{fig: main2} (a)) to the region $D$ only. 
Fig.\ref{fig: main2} (b) and (c) show the spin moment along $x$-axis, $S_x$ as a functions of time for the region $D$ and $D'$, respectively. Here, the dashed curves correspond to the periodic case where the coupling is conventional power-law decaying, and hence negligible. In this case, the spins in the region $D$ evolve along the local field axis but not for the spins in $D'$.  However, in the Fibonacci chain, due to the anomalously strong long-distance coupling, the localized spin on $D'$ region is realigned along the $x$-axis, despite of the absence of the external field on its position. It turns out that strong correlation given by the anomalous indirect interaction leads to the manipulation of the local magnetic moments in a non-local way. This reveals the strong entanglement over a long-distance beyond mesoscopic scale \cite{lu2022measurement,wei2022sequential,PhysRevB.105.094422,vovrosh2022dynamical,koffel2012entanglement}.




\textit{\tcm{Discussion and Conclusion---}} In summary, we unveil the anomalous long-distance quantum coupling emergent in the quasicrystals in terms of their inherent hyperspace geometry. The indirect interaction mediated by the critical states of quasicrystals show anomalous enhancement for an arbitrarily long distance. To analyze this phenomena, we introduce the concept of distance in the perpendicular space, which we call a hyperspace geometric distance. Based on this, we elucidate that the indirect interaction between spins on the quasiperiodic chains could be characterized by the hyperspace geometric distance instead of the physical distance between spins. The interaction universally decays as a function of hyperspace geometric distance. Therefore, the predominant interaction occurs between spins located on sites whose long-range environments coincide, irrespective of their physical separation. Hence, one can generate the entanglement over long distances. Our study offers quasiperiodic order as a groundbreaking way to induce strong long-range quantum correlations.

It is worth to note that the length scale of such anomalous interaction could be enhanced. Since the projection onto the perpendicular space densely covers the compact window\cite{jagannathan2021fibonacci}, the hyperspace geometric distance could be arbitrarily small. Hence, one can achieve the strong interaction for longer length scales even in macroscopic scales. It is also important to note that the length scale of the strong coupling can be controlled via different tiling patterns or changing the Fermi levels as well.

Quantum couplings that operate over long-distance indeed give rise to new paradigms in exploring exotic phases of matters such as topological phases, quantum spin liquids and  superconductivity\cite{zhou2017quantum,savary2016quantum,kim2013long,pezze2017multipartite,gong2017entanglement}. For instance, our result would be generalized to different interactions such as Cooper pairing mediated by critical phonon modes.
The non-uniform but self-similar spin exchanges in quasiperiodic chain will induce significant modification in the Doniach phase diagram allowing partial Kondo hybridization as well\cite{doniach1977phase}. 
It could generate competing exchange interactions between the spins over long-distance. Thus, magnetic frustration between them could lead to fractionalized spinons that operate in long-range. It gives rise to new types of quantum spin liquids and fractionalized excitations, which we leave as an interesting future work. Also, it would be worthwhile to investigate the anomalous long-distance interaction that emerge in heterostructures and higher dimensional systems \cite{PhysRevB.99.165430,mohammadi2015rkky,gonzalez2017electrically,sboychakov2018externally,PhysRevB.106.174436,vahedi2021magnetism,matsuo2005ising,ishikawa2016composition,thiem2015long,thiem2015magnetism,takeuchi2023high,suzuki2021magnetism}.

\section*{Acknowledgments}

J.M.J and S.B.L. were supported by National Research Foundation Grant (2021R1A2C109306013) and Nano Material Technology Development Program through the National Research Foundation of Korea(NRF) funded by Ministry of Science and ICT (RS-2023-00281839). This research was supported in part by grant NSF PHY-1748958 to the Kavli Institute for Theoretical Physics (KITP).

\bibliography{my1}
\bibliographystyle{unsrt}

\newpage

\renewcommand{\thesection}{\arabic{section}}
\setcounter{section}{0}
\renewcommand{\thefigure}{S\arabic{figure}}
\setcounter{figure}{0}
\renewcommand{\theequation}{S\arabic{equation}}
\setcounter{equation}{0}

\begin{widetext}
\section{Supplementary Texts}
\subsection{Review: Critical states in the Fibonacci chain}
	\label{sec:0}
Here, we briefly review the well-known properties of the critical states in the Fibonacci chain. We consider $H_T$ in the main text, the tight binding model with nearest neighbor hopping, which exhibit particle-hole symmetry, resulting in a symmetric spectrum. Fig.\ref{fig: supp0} (a) displays the spectrum of $H_T$ in the Fibonacci chain for $\rho=t_A/t_B=0.3$ as the function of the integrated density of states (IDOS). Notably, the spectrum is itself self-similar. This implies that the eigenstates are critical states characterized by the fractal wave functions. Indeed, almost all eigenstates of $H_T$ in the Fibonacci chain are critical, with the exception of a few localized states, such as $X$ and $Y$, which are marked in the inset of Fig.\ref{fig: supp0} (a).
\begin{figure}[h]
  \includegraphics[width=0.9\textwidth]{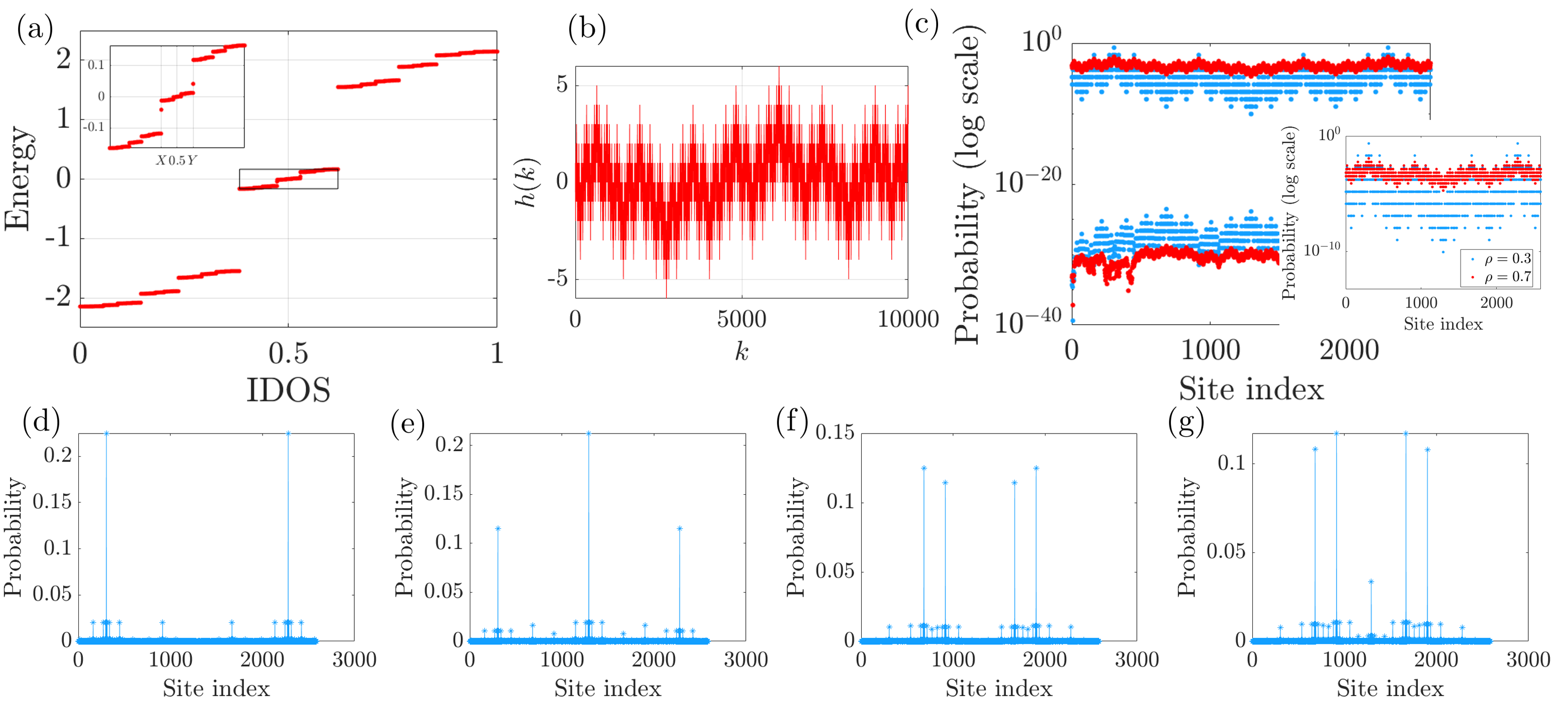}
  \caption{(a) Spectrum of the tight binding Hamiltonian with nearest neighbor hopping in the Fibonacci chain for $\rho=0.3$. The inset highlights the fractal nature of the spectrum, indicating exceptional localized modes marked by. $X$ and $Y$. (b) The height field of the Fibonacci chain. (c) Zero energy eigenstates of $H_T$ for $\rho=0.3$ and $0.7$, represented in blue and red, respectively. As $\rho$ deviates from 1, the wave function is more localized on a few sites that are sharing same local patterns. (d-g) Examples of the critical states near the zero energy, all calculated for $\rho=0.3$, $N=2585$. (d) 1293 (zero energy) (e) 1292 (f) 1291 (g) 1290-th states.}
  \label{fig: supp0}
\end{figure}

Fig.\ref{fig: supp0} (b) shows the height field of the Fibonacci chain. The characteristics of the height field such as fractal and scaling behavior are pattern-dependent characteristics. Specifically, the height field of the Fibonacci chain exhibits logarithmic growing, while different tiling patterns would have the linearly growing height field. Importantly, unless the zero mode is forbidden due to the gap labeling theorem, the wave function of the zero mode is given in terms of the height field and $\rho$. Thus, the localization characteristics of the zero energy state are also pattern dependent. See Refs.\onlinecite{kellendonk2015mathematics,PhysRevResearch.3.013168,PhysRevB.106.134431} for detailed information of the topological robustness of the height field for various aperiodic tiling patterns.

Fig.\ref{fig: supp0} (c) compares the zero energy states for different $\rho$ values. As $\rho$ deviates from 1, the critical state at zero energy is more concentrated on a few sites. We emphasize that this feature is general for the other critical states. For instance, we explicitly plot the critical states near the zero energy in Figs.\ref{fig: supp0} (d-g). For sufficiently strong quasiperiodicity (e.g. $\rho=0.3$), the critical states near the zero energy are concentrated on a few sites.

\subsection{Review: Detailed derivation of indirect exchange coupling}
Let us consider two localized magnetic moments at $i$ and $j$ sites, $\vec{S}_{i(j)}$ that are coupled with the itinerant electrons via local exchange coupling. The local exchange coupling term of the Hamiltonian is given by
\begin{align}
\label{KondoH}
&H_K=J_K(\vec{S}_i\cdot\vec{s}_i+\vec{S}_j\cdot\vec{s}_j)
\end{align}
Then, the indirect coupling is obtained by integrating out the itinerant electron degrees of freedom. In detail, for the effective Hamiltonian $\mathcal{H}=J_{ij}\vec{S}_i\cdot\vec{S}_j$, the coupling constant $J_{ij}$ is given by
\begin{align}
\label{couplingconst}
&J_{ij}=\frac{J_K^2}{4}\chi_{ij}.
\end{align}
Here, $\chi_{ij}$ is the susceptibility. From the linear response theory, the susceptibility is given by
\begin{align}
\label{susceptability}
&\chi_{ij}=-\frac{2}{\pi}\int_{-\infty}^{E_F}\mathcal{I}[G_{ij}^r(E)G_{ji}^r(E)]dE
\end{align}
where $G^r(E)$ is the retarded Green's function for the electronic tight binding Hamiltonian. $E_F$ is the Fermi energy. Then, one can write the Green's function in terms of the energy eigenstates of the tight binding Hamiltonian, $G_{ij}^r(E)=\lim_{\eta\to 0}\sum_m\frac{\psi_m(i)\psi_m(j)^*}{E-E_m+\mathrm{i}\eta}$. Here, $m$ is the index of the eigenstates, $E_m$ is the $m$-th eiegenenergy of the electronic spectrum. Since we are taking account of the imaginary part, we can rewrite Eq.\eqref{susceptability} as
\begin{align}
\label{susceptability2}
&\chi_{ij}=-\frac{2}{\pi}\int_{-\infty}^{E_F}\lim_{\eta\to0}\sum_{m,n}\left[\frac{\mathcal{R}[\psi_m(i)\psi_m(j)^*\psi_{n}(j)\psi_{n}^*(i)]}{\mathcal{I}[(E-E_m+\mathrm{i}\eta)(E-E_{n}+\mathrm{i}\eta)]}  +\frac{\mathcal{I}[\psi_m(i)\psi_m(j)^*\psi_{n}(j)\psi_{n}^*(i)]}{\mathcal{R}[(E-E_m+\mathrm{i}\eta)(E-E_{n}+\mathrm{i}\eta)]}\right]dE
\end{align}
However, since the second term of the integrand in Eq.\eqref{susceptability2} is odd under the exchange of the eigenstates indices, $m$ and $n$, it vanishes. Also, from the Sokhotski–Plemelj theorem, $\lim_{\eta\to0}(x+\mathrm{i}\eta)^{-1}=\mathcal{P}(1/x)-\mathrm{i}\pi\delta(x)$, we can get
\begin{align}
\label{susceptability3}
&\chi_{ij}=2\int_{-\infty}^{E_F}\sum_{m,n}\mathcal{R}[\psi_m(i)\psi_m(j)^*\psi_{n}(j)\psi_{n}^*(i)]\left( \frac{\delta(E-E_{n})}{E-E_m} +\frac{\delta(E-E_{m})}{E-E_{n}}  \right)dE
\end{align}
Note that Eq.\eqref{susceptability3} becomes nonzero only if $E_{m}<E_F$ and $E_n\ge E_F$ or vice versa. If the temperature is nonzero, then we can use the Fermi-Dirac distribution, $n_F(\varepsilon)=(1+\mathrm{e}^{\beta\varepsilon})^{-1}$ as shown in the main text.

\section{Supplementary Figures}
\subsection{Anomalous long-distance coupling in the general quasiperiodic tilings}
Let us extend our analysis to other quasiperiodic chain cases.
The anomalous couplings, which are strongly enhanced at long distances, are expected to appear in other quasiperiodic chains as well. To clarify this, let us consider the generalization of the Fibonacci substitution maps of two prototiles, that is $\mathrm{A}\to \mathrm{A}^r\mathrm{B}$ and $\mathrm{B}\to \mathrm{A}$. The resulting tilings are known as the metallic mean tilings. Note that the Fibonacci quasicrystal is the special case with $r=1$. For instance, the $r=2$ corresponds to the silver mean tiling. Also, the metallic mean tilings could be obtained by the cut-and-project scheme similar to the Fibonacci chain. In detail, the cut-and-project scheme with the physical space whose slope is inverse of metallic mean results in the corresponding metallic mean tiling with some proper hyperspace translations\cite{jagannathan2021fibonacci}. For these tilings, we may have critical states near the zero energy similar to the case of Fibonacci chain. This is because the height field defined on these tilings exhibits behaviors similar to those of the Fibonacci chain, including oscillations and logarithmic growth of its envelope\cite{PhysRevResearch.3.013168}.

We investigate the long-range interaction on these metallic mean tilings in terms of the real-space distance and hyperspace geometric distance. We set $E_F=0$. To define the hyperspace geometric distance, $D_{\pi^\perp}^M(i,j)$, we should determine the set of reference sites $M$, which depends on the specific tiling pattern. Remind that the set $M$ is comprised of the sites which have the largest probability amplitude of the zero energy state. For $\rho<1$, we find $\pi^\perp(M)=\{0\}$ for the silver mean tiling, but $\pi^\perp(M)=\{ \pm 0.4102, \pm 0.4360, \pm 0.4954, \pm0.5212 \}$ for the bronze mean tiling. Note that as $r$ increases, multiple locations in the perpendicular space correspond to similar long-range local tiling patterns due to the emergence of longer sequences of successive $\mathrm{A}$s, resulting in redundant images in the case of the bronze mean tiling. See Fig.\ref{fig: suppFSB} for detailed information of the set of the reference sites $M$ depending on the tiling pattern.
\begin{figure}[h]
  \includegraphics[width=0.6\textwidth]{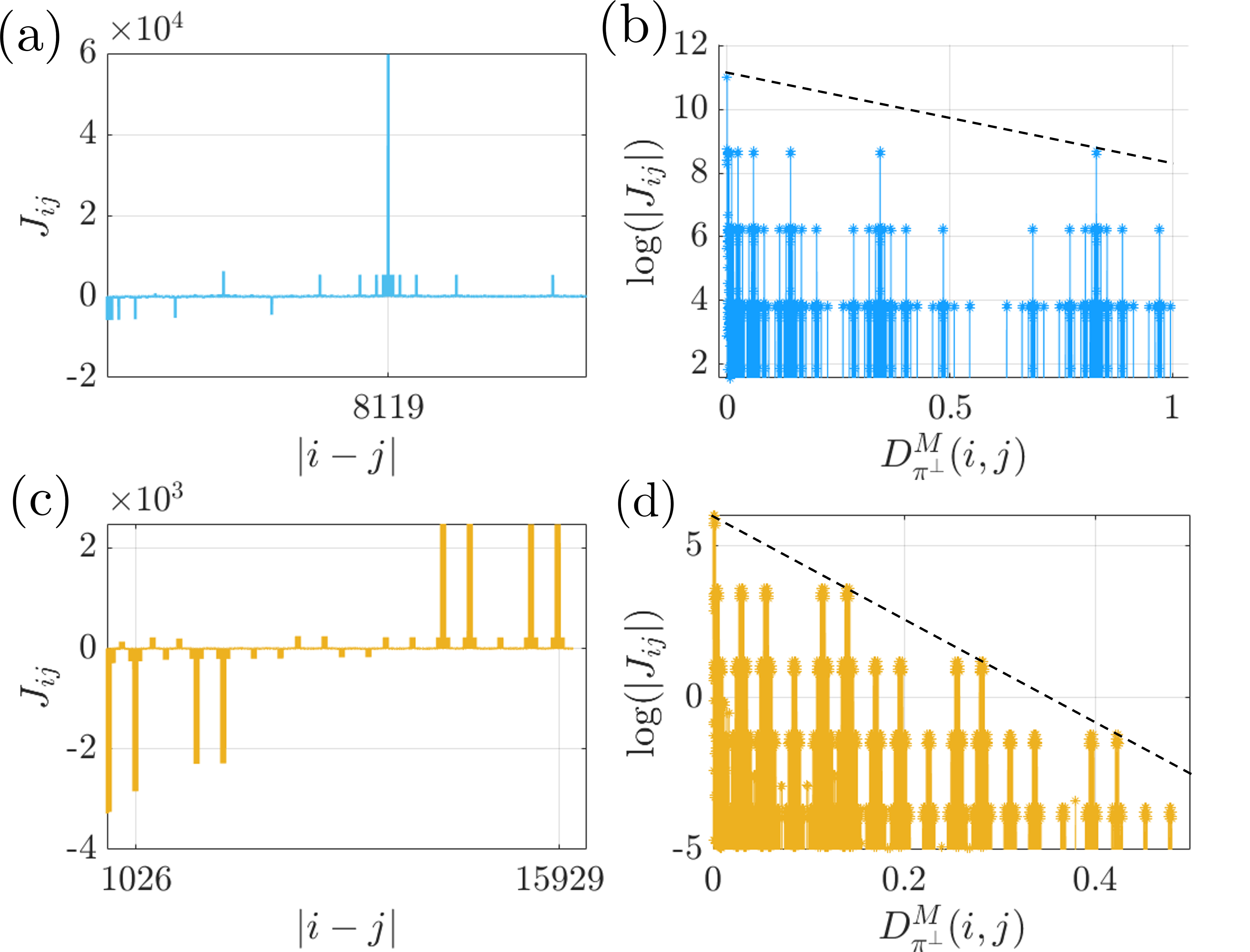}
  \caption{Typical long-range couplings $J_{ij}$ for general tiling cases with $\rho=t_A/t_B=0.3$ and $J_K=1$. (a,b) Silver mean tiling for $i=5741$ and (c,d) bronze mean tiling for $i=483$. The system sizes are (a,b) 19602, (c,d) 16868, respectively. (a,c) $J_{ij}$ as the function of the real-space distance, $|i-j|$. $J_{ij}$ are strongly enhanced for the mesoscopic length scales, $\mathcal{O}(10^4 a)$, where $a$ is the average lattice parameter. (b,d) $J_{ij}$ as the function of the hyperspace geometric distance, $D_{\pi^\perp}^M(i,j)$. For each case, $d_{\pi^\perp}(i,M)\approx 0$. Black dashed lines are included to highlight the overall decaying trends of the coupling strength with $D_{\pi^\perp}^M(i,j)$.}
  \label{fig: silverbronze}
\end{figure}

Figs.\ref{fig: silverbronze} (a) and (c) display examples of anomalous long-range couplings in the silver ($r=2$) and bronze ($r=3$) mean tilings as functions of the physical distance. Importantly, Fig.\ref{fig: silverbronze} (a) and (c) demonstrate that the anomalously enhanced long-range couplings observed beyond the mesoscopic scale are not limited to any specific tiling pattern; instead, they are a characteristic feature of quasiperiodic chains in general. On the other hand, Fig.\ref{fig: silverbronze} (b) and (d) illustrate the $J_{ij}$ as the function of hyperspace geometric distance. Here, we select $i$ sites as $d_{\pi^\perp}(i,M)\approx 0$. The most dominant interaction arises at the shortest hyperspace geometric distance. In quasiperiodic chains, the primary interactions occur between sites that are characterized by a certain similar long-range tiling pattern in their local environment. Thus, the hyperspace geometric distance serves as a better indicator for analyzing dominant long-range interactions in general quasiperiodic chains than physical distance. The black dashed curves in Figs.\ref{fig: silverbronze} (b) and (d) emphasize that the overall coupling strength decreases as hyperspace geometric distance increases. However, comparing Figs.\ref{fig: silverbronze} (b,d) and Fig.2 (e) in the main text, we observe that the decay of the coupling strength exhibits some plateaus for silver and bronze mean cases. This phenomenon arises because the perpendicular space contains numerous redundant regions that correspond to locally isomorphic patterns as $r$ increases. See Fig.\ref{fig: supplesilb} for detailed analysis of the silver ($r=2$) and bronze ($r=3$) mean cases.

The length scales of dominant long-range couplings are influenced by both the tiling pattern and system size. As $r$ increases, the predominant interaction length scale generally expands. This expansion is attributed to the quasiperiodic tiling pattern manifesting over longer length scales with increasing $r$. Moreover, in the thermodynamic limit, interactions at arbitrarily long length scales can be observed, even at the macroscopic level, within a specific quasiperiodic tiling pattern. This arises from the dense nature of the $\pi^\perp$ projection image, which results in no lower bound on hyperspace geometric distance. Specifically, for the Fibonacci chain, the main interaction length scale is approximately $\sim \mathcal{O}(10^3 a)$, where $a$ represents the average lattice parameter. In contrast, for larger system sizes, the silver mean and bronze mean tilings exhibit longer length scales, approximately on the order of $\sim \mathcal{O}(10^4 a)$, as shown in Figs.\ref{fig: silverbronze} (a) and (c).

\subsection{Concentration of the wave function near the zero energy}
In this section, we discuss the concentration of the wave function near the zero energy in various tilings. First, let us compare the product of the wave functions, $\vert\psi_n(i)\psi_n(j)\vert$ for different $j$ sites in periodic chain to explain the universal decaying with distance. Let us consider $i=17$ for concrete argument, however, we emphasize that our analysis is generally applicable for the other sites. In the periodic chain, the energy eigenstates are Bloch wave functions, and hence the position eigenstates do not lie on the specific energy levels. Fig.\ref{fig: periodicanalysis} (a) and (b) show the amplitude of the product of the wave functions, $\vert\psi_n(i)\psi_n(j)\vert$, where $j=19$ and $j=1292$, respectively. Remarkably, as the distance between two sites increases, $\vert\psi_n(i)\psi_n(j)\vert$ rapidly oscillates near the zero energy (compare Fig.\ref{fig: periodicanalysis} (c) and (d)). Furthermore, $\vert\psi_n(i)\psi_n(j)\vert$ is not concentrated near the zero energy. Since the contribution to the long-range coupling becomes negligible as the energy is far away from the Fermi level (see Eq.\eqref{susceptability3}), the interaction strength in the periodic chain decays uniformly with distance.
\begin{figure}[h]
  \includegraphics[width=0.5\textwidth]{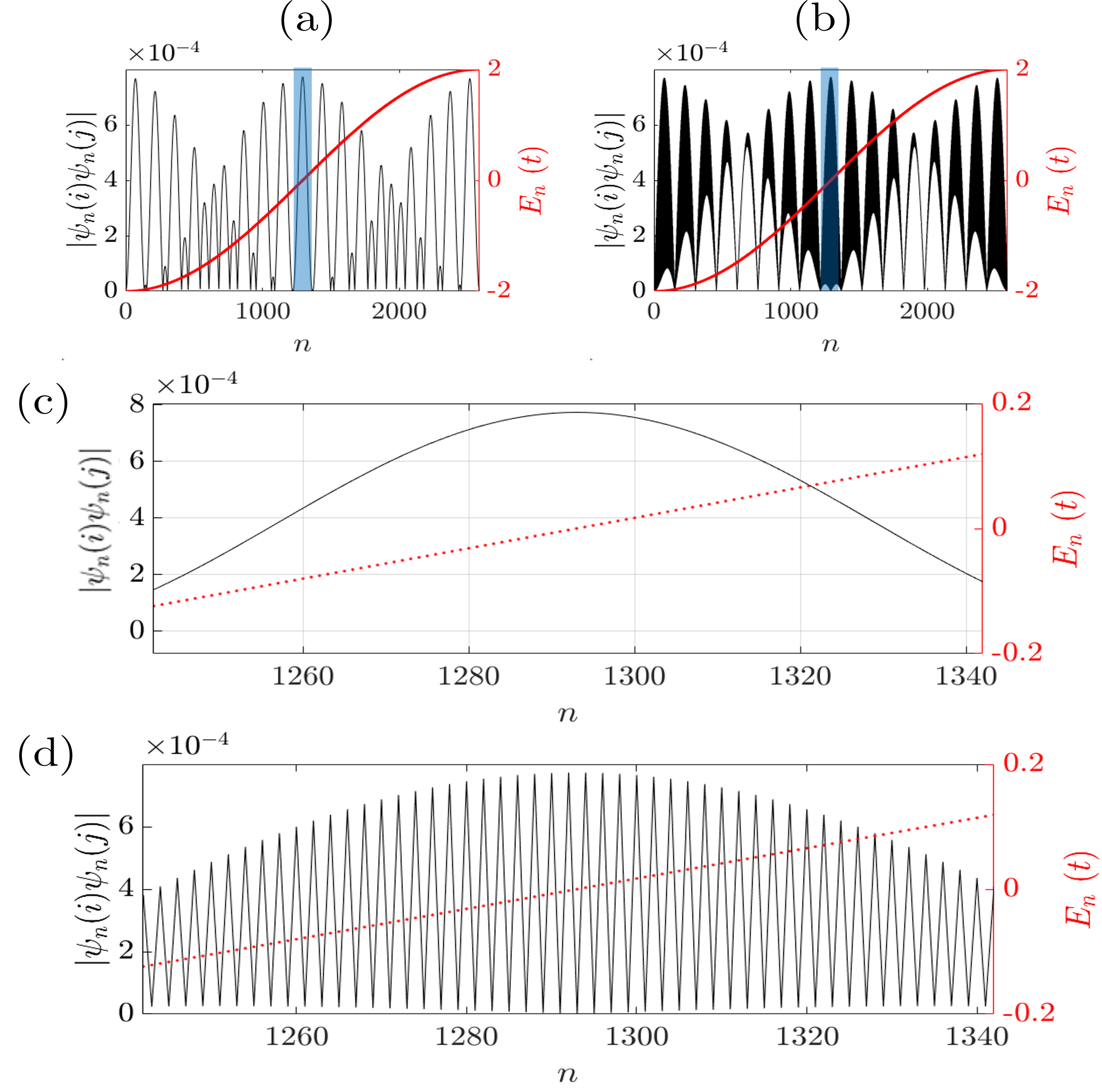}
  \caption{Energy eigenstate analysis for the periodic chain. Amplitude of the product of the wave functions, $\vert\psi_n(i)\psi_n(j)\vert$ for (a) $j=19$ and (b) $j=1292$ as the function of energy index $n$. The blue shaded region indicates that the near zero energy. (c) and (d) are the zoom-in distribution of the blue shaded region in (a) and (b), respectively. The system size is 2585, $t=1$ and $i=17$.}
  \label{fig: periodicanalysis}
\end{figure}

Now let us consider the Fibonacci pattern generated by $\rho\neq 1$. To explain how $J_{ij}$ is enhanced for a long distance with the quasiperiodicity, we compare the wave functions and energy spectrum for different $\rho$ values. Fig.\ref{fig: addmain} displays the wave function, $\psi_n(i)$ as a function of $n$ for a given $i$ site along with the energy spectrum. Here, we consider $i=17$ and $j=1292$ as a particular example showing anomalous coupling in the Fibonacci chain. Since the main contributions to the long-range exchange coupling are originated from the states near the Fermi level, $\psi_n(i)\psi_n(j)$ with $E_n\approx E_F=0$ is dominant in $J_{ij}$. Note that in the Fibonacci chain, the position state can only occupy a limited number of energy levels, as most states are critical. Thus, $\psi_n(i)$ is concentrated on a few $n$ values (see Fig.\ref{fig: addmain} (g)). Particularly, since the critical wave function is concentrated on the multiple sites that share the same local environment, there are numerous pairs of sites $(i,j)$ whose $\vert\psi_n(i)\psi_n(j)\vert$ is also more concentrated and exhibits less oscillation near zero energy compared to the uniform chain case, regardless of the distance $\vert i-j\vert$ (see Fig.\ref{fig: addmain} (h) and (i).). Thus, in the Fibonacci chain, there are many pairs of sites that, despite being far apart, occupy the same energy levels near the zero energy and are generally strongly correlated with each other. As a result, the interactions between these pairs of sites are enhanced, even over long distances. In addition, in the Fibonacci chain, the energy spectrum becomes almost flat near the zero energy, opening many tiny gaps (compare the energy levels (red color, right $y$-axis) in (a-c) and (g-i)). Since this reduces the denominator of Eq.\eqref{susceptability3}, $J_{ij}$ is generically enhanced in the Fibonacci chain.

Comparing Figs.\ref{fig: addmain} (d-f) and (g-i) for $\rho=0.7$ and $0.3$, it also turns out that as $\rho$ deviates from 1, near the zero energy, the spectrum becomes more flat and $\vert\psi_n(i)\psi_n(j)\vert$ is more concentrated. Hence, the nature of the anomalous long-distance interaction becomes more pronounced as the quasi-periodicity becomes stronger. This indicates the manageability of the long-distance strong coupling in terms of the strength of the quasiperiodicity, $\rho$. Explicitly, Fig.\ref{fig: addmain} (j) shows $\rho$-dependence of the long-range interactions, $J_{ij}$. As $\rho$ deviates from 1, the anomalous long-range couplings are monotonically enhanced. Particularly, for sufficiently strong quasiperiodicity, such as $\rho<0.6$, long-distant sites can exhibit much stronger interactions than the nearest neighbor interaction $J_{\mathrm{n.n}}$ for the given $\rho$.
\begin{figure*}
  \includegraphics[width=1.0\textwidth]{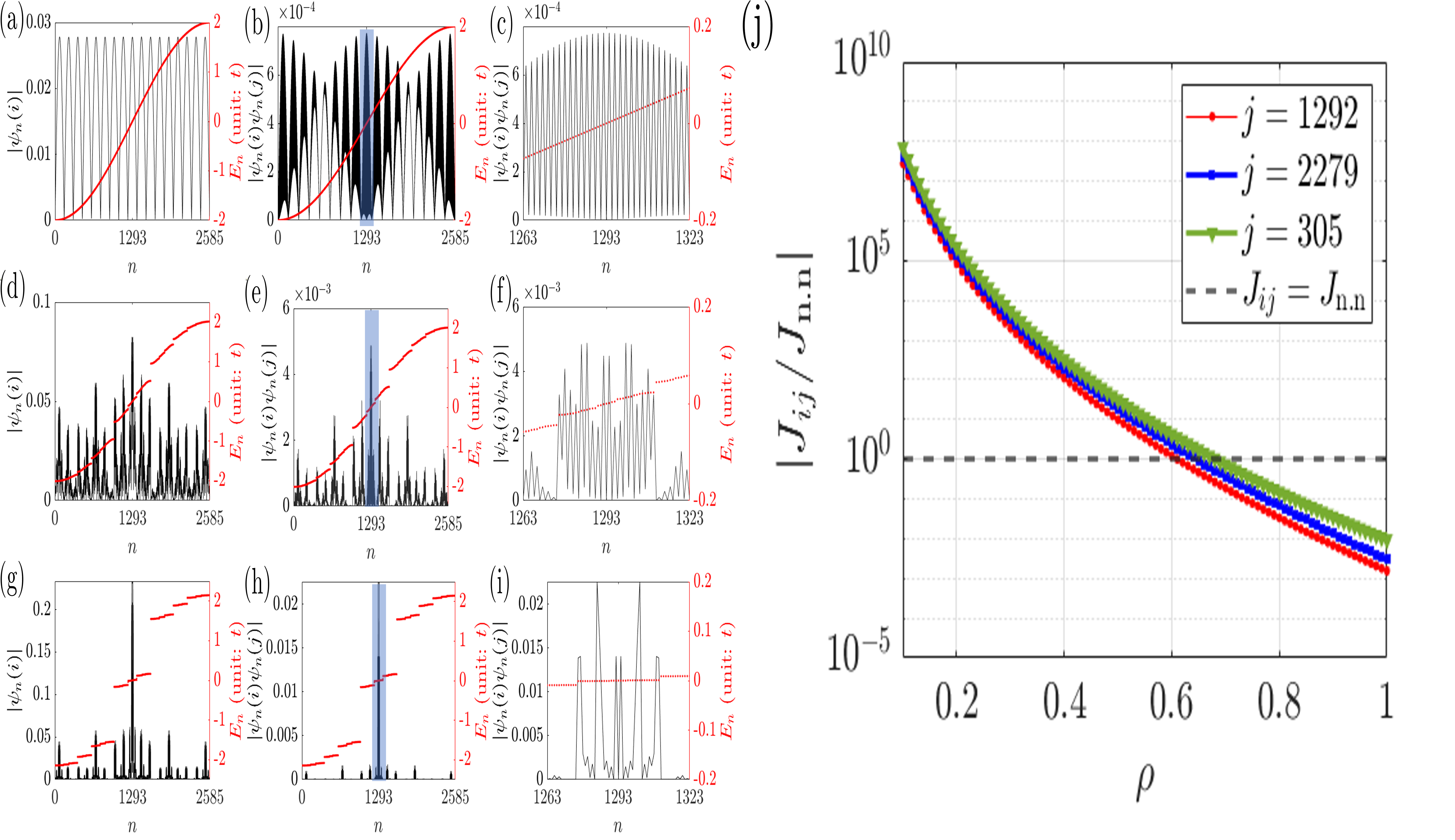}
  \caption{Analysis based on the wave function amplitude $\psi_n(i)$ as a function of energy index $n$ for different hopping ratios, $\rho=t_A/t_B$. (a-c) $\rho=1$, periodic chain. (d-f) $\rho=0.7$ and (g-i) $\rho=0.3$. Here, $i=17$ and $j=1292$. (a,d,g) $\vert\psi_n(i)\vert$ as a function of $n$.(b,e,h) $\vert\psi_n(i)\psi_n(j)\vert$ as a function of $n$. Blue shaded region is drawn for emphasizing the $n$ values near the zero energy that mainly contribute to the long-range coupling. (c,f,i) Zoom-in view of the blue shaded region in (b), (e) and (h), respectively. For comparison, energy spectrum is also drawn (red color, right $y$ axis for each panels). (j) For given $i=17$ and various $j=305$ (green triangle), $1292$ (red circle), $2279$ (blue square), the relative strength of $J_{ij}$ and nearest neighbor interaction, $J_{n.n}$ as the function of $\rho$. The system size $N=2585$. $J_K=1$ and $t=1$.}
  \label{fig: addmain}
\end{figure*}

We emphasize that the enhancement of $\vert J_{ij}/J_\mathrm{n.n}\vert$ in Fig.\ref{fig: addmain} (j) arises from both the decrease of $\vert J_\mathrm{n.n}\vert$ and the increase of $\vert J_{ij}\vert$. To show this, we examine $\vert \frac{J_{ij}}{J_\mathrm{n.n}(\rho=1)}\vert$. Fig.\ref{fig: supp_ref_C1} shows that $\vert \frac{J_{ij}}{J_\mathrm{n.n}(\rho=1)}\vert$ monotonically increases as $\rho$ deviates from 1. Thus, the long-range interaction for $\rho\neq 1$ is indeed enhanced compared to the short-range interaction in the $\rho = 1$ (uniform) case. We also note that as $\rho$ moves away from 1, the short-range interaction diminishes. Therefore, the strongly correlated phenomena associated with strong quasiperiodicity are driven by the anomalously enhanced long-range interactions.
\begin{figure*}
  \includegraphics[width=0.5\textwidth]{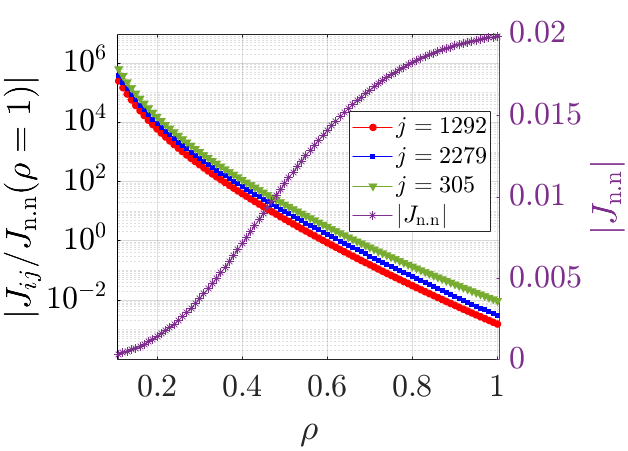}
  \caption{For a given $i=17$ and various sites $j$, including $j=305$ (green triangle), $1292$ (red circle), $2279$ (blue square), the relative strength of $J_{ij}$ is presented compared to the nearest neighbor interaction, $J_{n.n}$ for $\rho=1$, as a function of $\rho$. The violet curve (right $y$-axis) shows that the short-range interaction, $J_\mathrm{n.n}$ diminishes as $\rho$ moves away from 1. The system size $N=2585$. $J_K=1$ and $t=1$.}
  \label{fig: supp_ref_C1}
\end{figure*}

Similar results can be found in the silver mean and bronze mean tilings. Fig.\ref{fig: supplesilb} illustrates the cases of the silver mean (see Fig.\ref{fig: supplesilb} (a-c)) and the bronze mean tilings (see Fig.\ref{fig: supplesilb} (d-f)). These quasiperiodic chains also possess position eigenstates with energies highly concentrated near the Fermi level, similar to the case of the Fibonacci chain. Here, we use $\rho=0.3$ for both tilings. Specifically, the sites coupled through long-range strong interactions show a highly concentrated product of amplitudes. $\vert\psi_n(i)\psi_n(j)\vert$, near the Fermi level, $E_F=0$.
\begin{figure}[h]
  \includegraphics[width=0.65\textwidth]{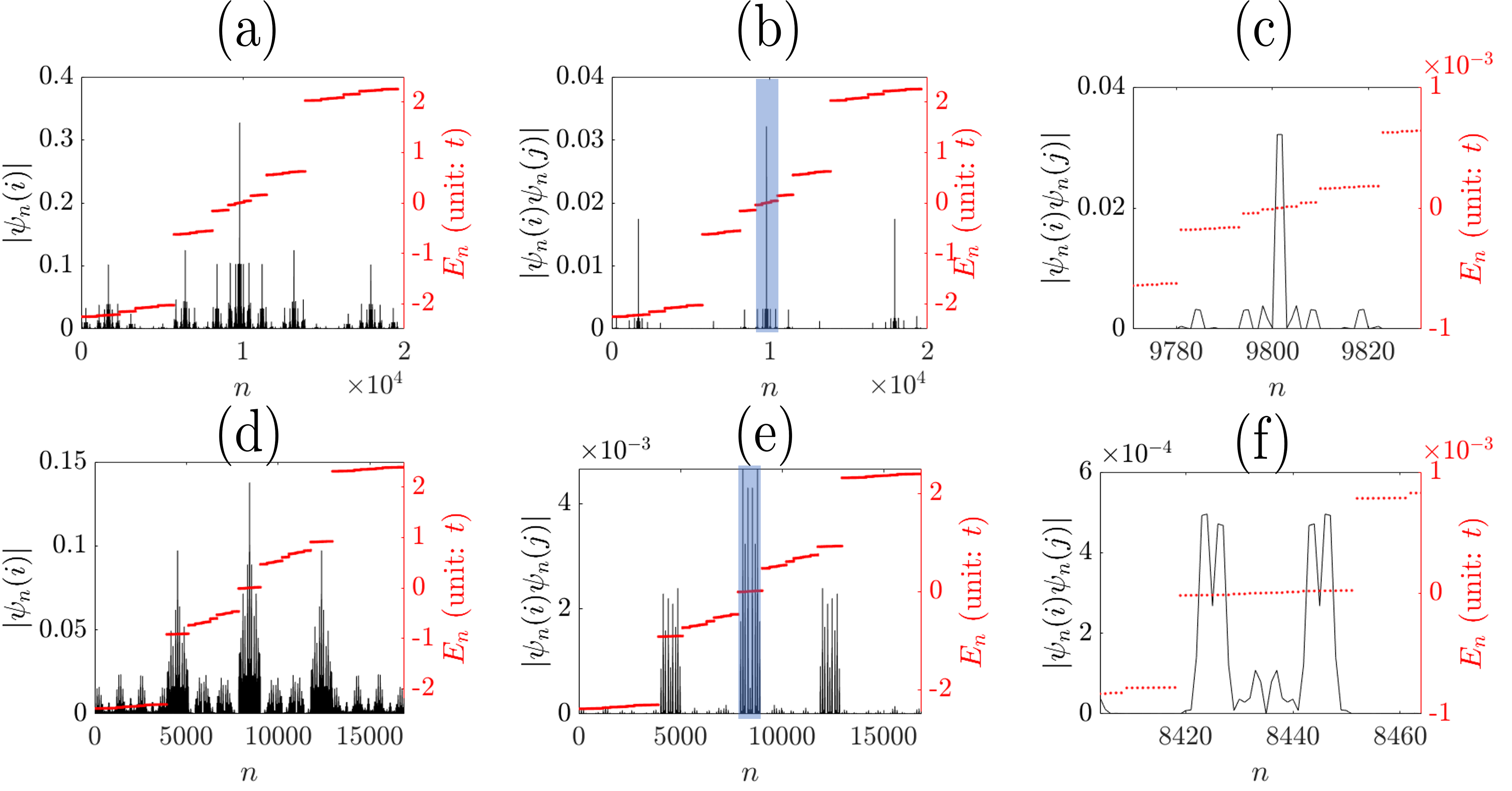}
  \caption{Energy eigenstate analysis for the silver mean (a-c) and bronze mean (d-f) tiling cases. (a) Energy eigenstate decomposition of the position eigenstate $i=5741$ in the silver mean tiling. (b) Amplitude of the product of the wave functions, $\vert\psi_n(i)\psi_n(j)\vert$ for $j=13860$ as the function of energy index $n$. (c) A zoomed-in view near zero energy that is emphasized as blue shaded region in (b). (d) Energy eigenstate decomposition of the position eigenstate for $i=483$ in the bronze mean tiling. (e) Amplitude of the product of the wave functions, $\vert\psi_n(i)\psi_n(j)\vert$ for $j=16412$ as the function of energy index $n$. (f) A zoomed-in view near zero energy that is emphasized as blue shaded region in (e). The system sizes are 19602 and 16898 for silver mean and bronze mean tilings, respectively. $t=1$ and $\rho=t_A/t_B=0.3$.}
  \label{fig: supplesilb}
\end{figure}

\subsection{Long-range coupling in the periodic approximant}
\label{sec:1}
Fig.\ref{fig: supp2} illustrates the long-range coupling in the periodic approximant of the Fibonacci chain. Unlike the Fibonacci chain, the interaction decays uniformly in the periodic approximant, similar to the RKKY interaction in the conventional crystals. Specifically, the exceptional enhancement of the long-range coupling appears limited to an unit cell of the approximant. This is because the local quasiperiodic order appears in an unit cell of the periodic approximant only. Thus, the anomalously enhanced long-range interaction is an unique feature of the quasiperiodic system.
\begin{figure}[h]
  \includegraphics[width=0.65\textwidth]{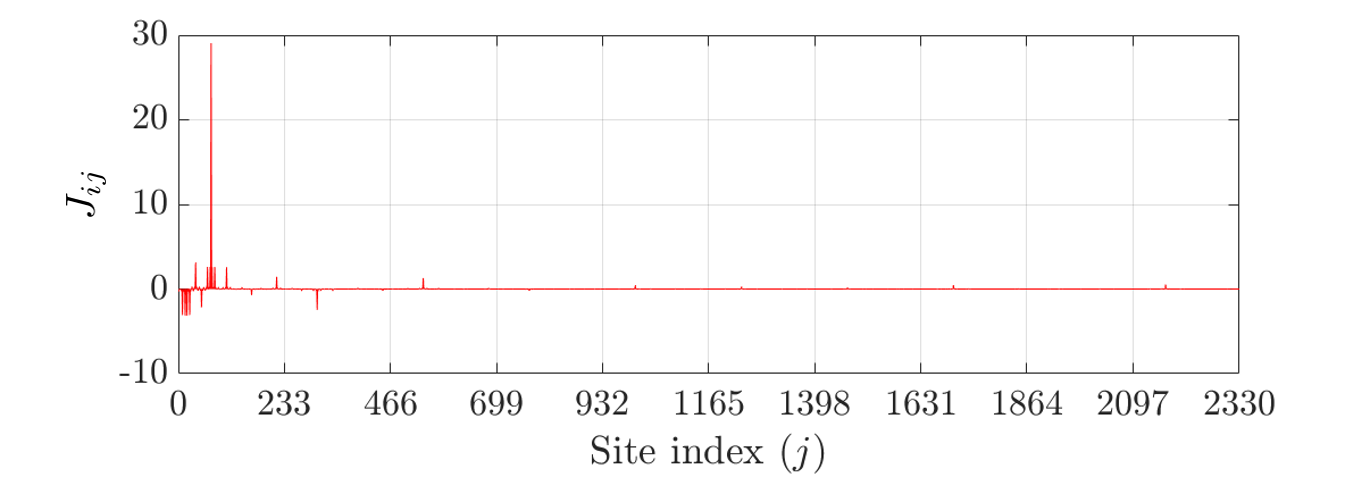}
  \caption{Typical long-range interaction, $J_{ij}$ in the periodic approximant of the Fibonacci chain whose unit cell contains 233 sites. Here, $i=17$. The system size is 2331. Anomalous characteristics of the long-range coupling in the approximant is appearing limited to each unit cells. The long-range coupling is universally decaying in between the unit cells of the approximant. Here, $E_F=0$, $\rho=t_A/t_B=0.3$ and $J_K=1$.}
  \label{fig: supp2}
\end{figure}

On the other hand, we emphasize that the anomalous long-range interaction is not significantly affected by changes in the boundary conditions. Comparing Fig.2 (a) in the main text, Fig. \ref{fig: supp_refC2A} shows a similar anomalous enhancement of the long-range couplings for the $i=17$ site in the Fibonacci chain with periodic boundary conditions as a function of distance. In this case, we impose the periodic boundary condition by identifying the left and right ends of the open-boundary Fibonacci chain. This suggests that the anomalous long-range interaction stems from the quasiperiodic bulk pattern rather than the boundary conditions.
\begin{figure}[h]
  \includegraphics[width=0.65\textwidth]{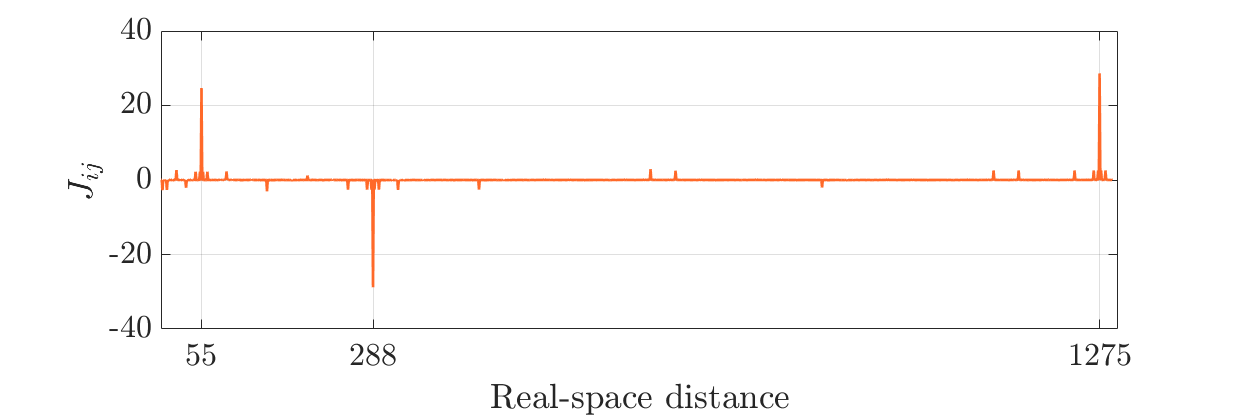}
  \caption{Anomalous long-range indirect interactions, $J_{ij}$ under the periodic boundary condition as a function of real-space distance. The system size is 2584 and $i=17$. $E_F=0$, $\rho=0.3$ and $J_K=1$.}
  \label{fig: supp_refC2A}
\end{figure}


\subsection{Conventional coupling in the Fibonacci chain}
Although most of sites have the anomalously enhanced interaction for the long-distance, there are some sites in the Fibonacci chain which admit the conventionally decaying interaction. Fig.\ref{fig: supp3} shows the coupling constant, $J_{ij}$ for the $i=16$ and $i=2278$ sites as the function of the site index $j$ in the Fibonacci chain. Here, we set $E_F=0$ and $\rho=0.3$. Note that $J_{ij}$ for these sites are not only rapidly decaying but also negligible compared to the anomalously enhanced couplings between $i=17$ and $i=2279$, for instance (See Fig. 2 in the main text for comparison.). As a result, the dynamics of the local magnetic moments placed on $i=16$ and $i=2278$ are conventional as shown in Fig.3 in the main text. Note that these sites are characterized by large hyperspace geometric distance.

\begin{figure}[h]
  \includegraphics[width=0.6\textwidth]{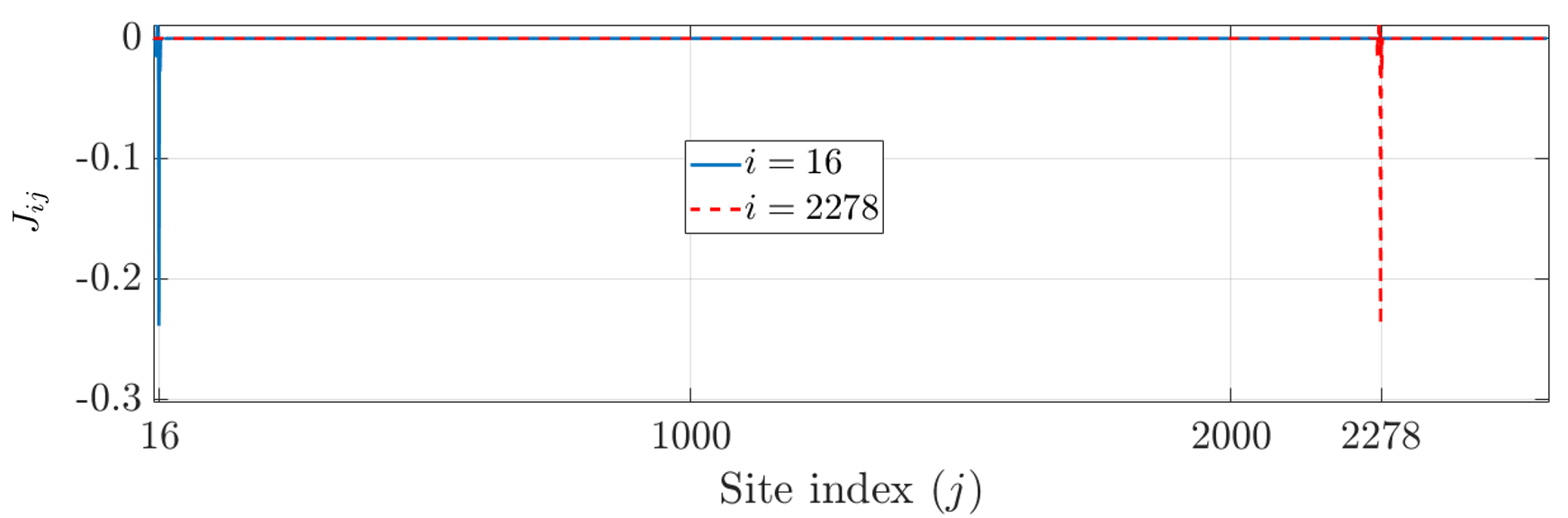}
  \caption{Indirect exchange coupling constant, $J_{ij}$ for $i=16$ (blue solid line) and $i=2278$ (red dashed line) sites as the functions of $j$. The coupling strengths for these sites decay rapidly with $\vert i-j\vert$. Here, $J_K=1$, $N=2585$, $\rho=0.3$ and $E_F=0$.}
  \label{fig: supp3}
\end{figure}

\subsection{Determination of the set of reference sites}
The set of reference sites, $M$ is depending on both the parameters of the model such as $\rho$ and tiling patterns. Assuming half-filled case with $E_F=0$, the set of reference sites could be obtained by the height field which is tiling pattern-dependent function. In detail, for $\rho<1$, the set of reference sites, $M$ is given by $M=\{ x\vert h(x)\le h(y) \ \ \forall y\in\pi \}$. Note that this set of reference sites is comprised of the positions where the zero energy critical state is maximally concentrating on. Even though their physical locations are unrevealing in the physical space, their corresponding perpendicular space projection images, $\pi^\perp(M)$ are informative. This is because the perpendicular space classify the physical positions in terms of their local surroundings.
\begin{figure}[h]
  \includegraphics[width=0.7\textwidth]{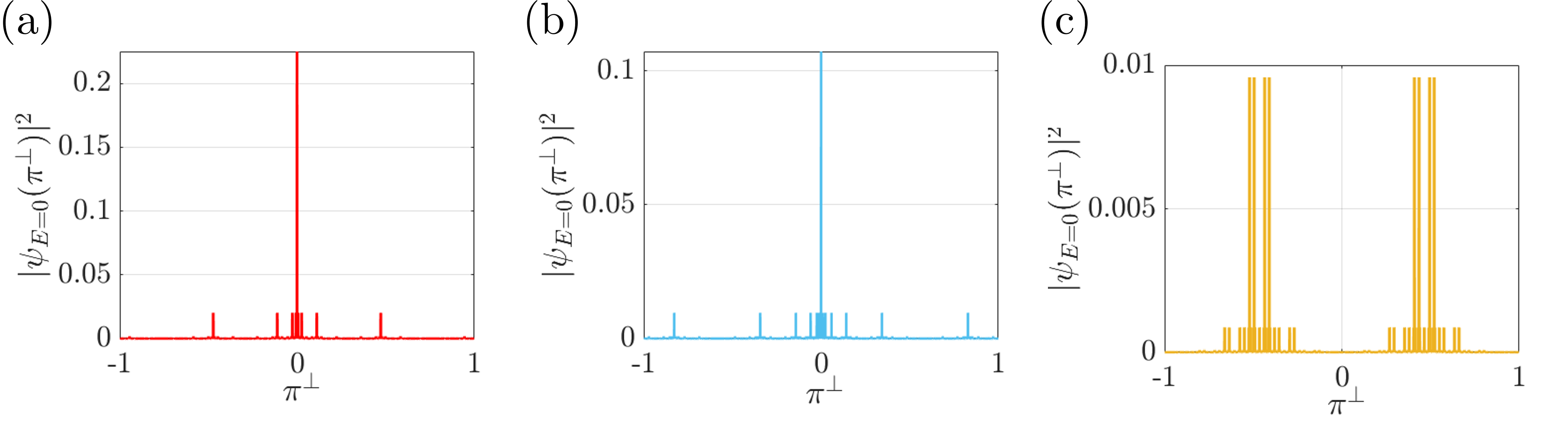}
  \caption{Probability distributions on the perpendicular space of the zero energy eigenstates of the different tiling patterns. (a) Fibonacci chain ($N=2585$) (b) silver mean ($N=19602$) (c) bronze mean ($N=16898$) tilings.  Here, $\rho=0.3$ and $t=1$. The zero energy state, $\psi_{E=0}$ is maximally concentrated on the zero of the $\pi^\perp$ space for the cases of the Fibonacci chain and silver mean tiling. On the other hand, the maximal value of the probability of the zero energy state appears multiple perpendicular space positions for the case of the bronze mean tiling. Thus, $\pi^\perp(M)=\{0\}$ for the Fibonacci chain and silver mean tiling, while $\pi^\perp(M)=\{ \pm 0.4102, \pm 0.4360, \pm 0.4954, \pm0.5212 \}$ for the bronze mean tiling.}
  \label{fig: suppFSB}
\end{figure}
Fig.\ref{fig: suppFSB} illustrates the $\pi^\perp$ space distribution of the zero energy critical states, $\psi_{E=0}$ for the Fibonacci, silver, and bronze mean tilings. Particularly, the bronze mean tiling case shows multiple positions where the critical state is maximally concentrated on. Thus, $\pi^\perp(M)$ contains multiple positions on the perpendicular space in the case of the bronze mean tiling. Specifically, $\pi^\perp(M)=\{ \pm 0.4102, \pm 0.4360, \pm 0.4954, \pm0.5212 \}$ for the bronze mean tiling. On the other hand, for the cases of the Fibonacci and silver mean tilings, $\pi^\perp(M)$ contains the center of the perpendicular space which is the origin of $\pi^\perp$ space only.

\subsection{Extension to General Fermi Levels}
In this section, we demonstrate that the anomalous long-range interaction and the decay of envelopes of the interaction strength with respect to the hyperspace geometric distance, are not limited to half-filling but are also applicable to more general Fermi levels. Let us consider the Fibonacci chain with a nonzero Fermi level, specifically $E_F=-1.909$, which is $\sim 20\%$ filling, as a concrete example to validate our two main claims at the general Fermi levels.

Fig.\ref{generalfilling2} (a) displays the energy spectrum of the Fibonacci chain with $\rho=0.3$ (in red) along with the Fermi level, $E_F=-1.909$ (in blue dashed line). Fig.\ref{generalfilling2} (b) shows $\vert\psi_{\varepsilon_F}\vert^2$ in the perpendicular space for the given $E_F$. Note that they are pair in $\pm|\pi^{\perp}|$ due to palindromic characteristics of the Fibonacci chain. The set $\pi^\perp(M)$ is defined by the perpendicular space positions of the largest peaks. Fig.\ref{generalfilling2} (c) demonstrates the presence of the anomalously enhanced long-range interaction for this general Fermi level. Meanwhile, Fig.\ref{generalfilling2}(d) illustrates how the long-range interaction decays as a function of the hyperspace geometric distance. Lastly, Fig.\ref{generalfilling2} (e) confirms that, similar to Fig.2 (e) in our main text, the envelope of the anomalous long-range interaction strength decreases with increasing hyperspace geometric distance.

In conclusion, the results in Fig.\ref{generalfilling2} indicate that the anomalous long-range interaction and the decay of interaction strength with hyperspace geometric distance are general features observable at any filling.
\begin{figure}[h]
  \includegraphics[width=0.5\textwidth]{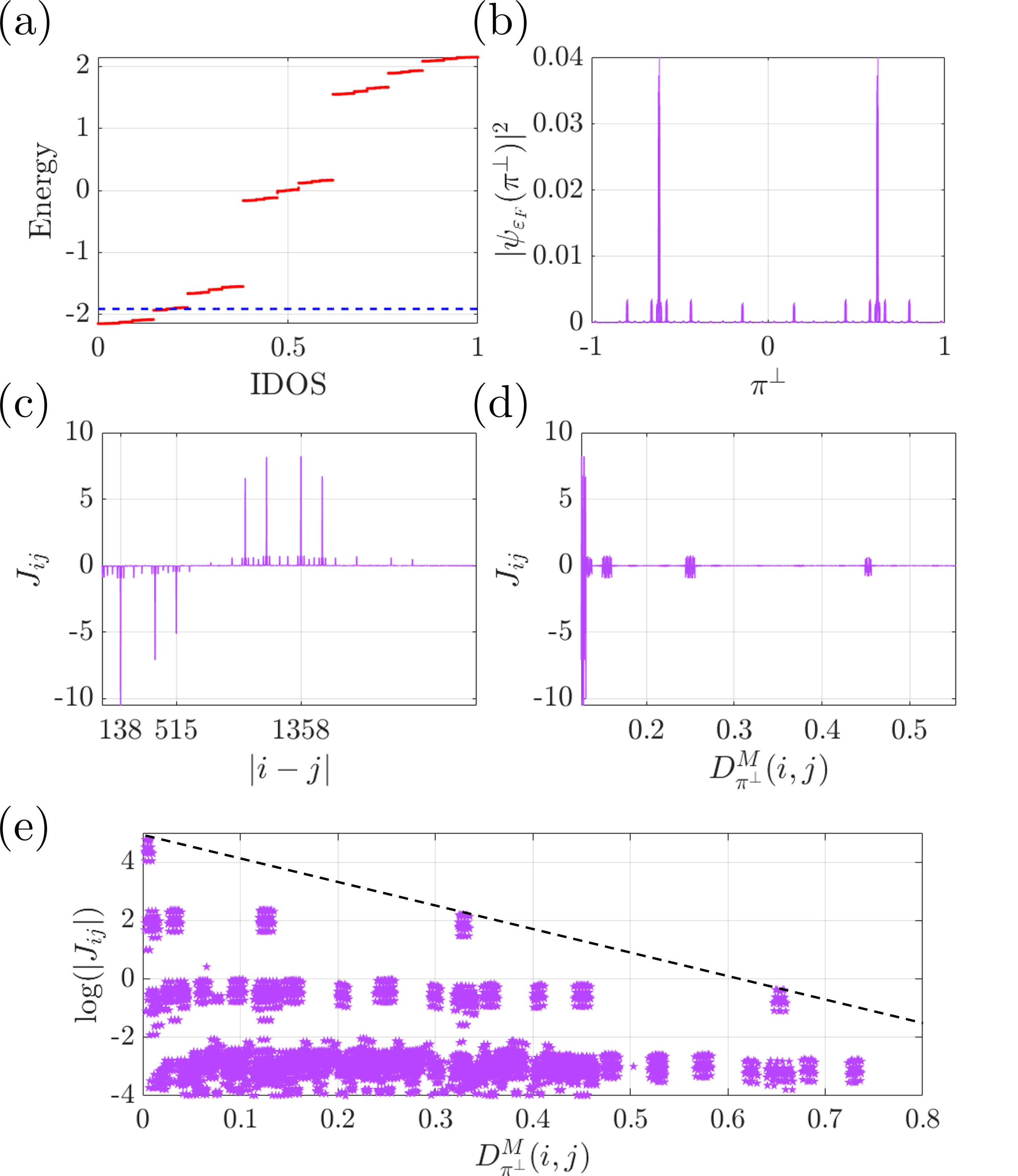}
  \caption{(a) The energy spectrum of the Fibonacci chain with $\rho=0.3$ (red) along with the chosen Fermi level $E_F=-1.909$ (blue dashed line). (b) The probability distribution $\vert \psi_{\varepsilon_F}\vert^2$ plotted in perpendicular space for the given $E_F$.  (c) Demonstration of the anomalously enhanced long-range indirect interaction observed at this Fermi level. Here, $i=50$. (d) The plot of (c) as a function of the hyperspace geometric distance. (e) The anomalous long-range interaction strength plotted against the hyperspace geometric distance for general pairs of sites, with black dashed line highlighting the exponential decay of the envelopes. The system size is 2585. $J_K=1$ and $t=1$.}
  \label{generalfilling2}
\end{figure}

Note that the set $M$ depends on the choice of the Fermi level. Fig.\ref{EFM2} shows the perpendicular space projection images of the set $M$, $\pi^\perp(M)$ as the function of $E_F$. 
\begin{figure}[h]
  \includegraphics[width=0.45\textwidth]{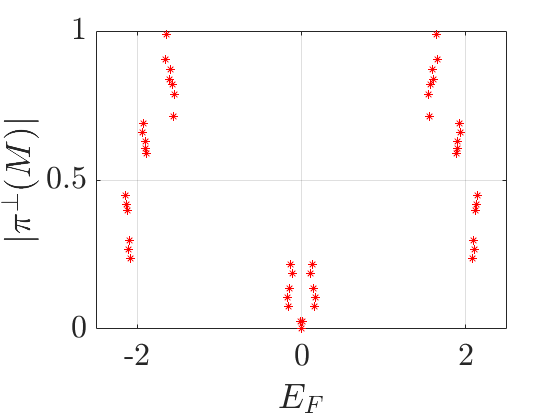}
  \caption{The perpendicular space projection images of the set of reference sites $M$ are shown as a function of the Fermi level, $E_F$ in the Fibonacci chain. The system size is 2585. $\rho=0.3$ and $t=1$.}
  \label{EFM2}
\end{figure}

\section{Methods}\label{secmethod}

\subsection*{LLG equation and long-distance spin manipulation}
The dynamics of the local magnetic moment at the $i$ site, $\vec{S}_i$ is governed by the Landau–Lifshitz–Gilbert equation,
\begin{align}
\label{LLG}
&\frac{d\vec{S}_i}{dt}=-\gamma \vec{S}_i\times \vec{H}_{\mbox{eff}}^i-\alpha\vec{S}_i \times(\vec{S}_i\times\vec{H}_{\mbox{eff}}^i),
\end{align}
where $\vec{H}_{\mbox{eff}}^i=-\nabla_{\vec{S}_i} \mathscr{H}$, $\alpha$ is a damping parameter. and $\gamma$ is the gyromagnetic ratio. Here, the Hamiltonian including both the long-range interaction and local magnetic field, $\mathscr{H}$ is given by,
\begin{align}
\label{HLLG}
&\mathscr{H}=\sum_{i\neq j} J_{ij} \vec{S}_i\cdot\vec{S}_j-\sum_{i\in D}\vec{S}_i\cdot \vec{B},
\end{align}
where $\vec{B}$ is the external magnetic field along $x$-axis applied to the local region, $D$. For numerical calculation shown in Fig.3 in the main text, we set $\alpha=0.1$ and $\gamma=1$.

\end{widetext}

\end{document}